\documentclass[aps,twocolumn,showpacs,preprintnumbers,amsmath,amssymb,
]{revtex4-2}
\usepackage{graphicx}
\usepackage{bm}
\usepackage{color}
\usepackage{units}
\usepackage{wasysym}
\usepackage{MnSymbol}
\usepackage[unicode=true,pdfusetitle,bookmarks=false,colorlinks=true,citecolor=blue,urlcolor=blue,linkcolor=blue]{hyperref}
\usepackage{caption}
\begin{document}

\title{A minimal model for liquid-liquid phase separation and aging of chemically reactive macromolecular mixtures}

\author{Ruoyao Zhang}
\affiliation{Department of Mechanical and Aerospace Engineering, Princeton University, Princeton, New Jersey 08544, United States}
\author{Sheng Mao}
\affiliation{Department of Mechanics and Engineering Science, BIC-ESAT, College of Engineering, Peking University, Beijing 100871, China}
\author{Mikko P. Haataja}
\email{mhaataja@princeton.edu}
\affiliation{Department of Mechanical and Aerospace Engineering, Princeton University, Princeton, New Jersey 08544, United States;}
\affiliation{Princeton Institute for Science and Technology of Materials (PRISM), Princeton University, Princeton, New Jersey 08544, United States}

\begin{abstract}
Mixtures of several macromolecular species can lead to the formation of higher-order structures that often display non-ideal mixing behavior. In this work, we propose a minimal model of a quaternary system which considers the formation of a complex via a chemical reaction involving two macromolecular species; the complex may then phase separate from the buffer and undergo a further transition into a gel-like state over time. First, a ternary phase diagram that captures the volume fraction of each species and phases at equilibrium is constructed. Specifically, we investigate how physical parameters such as stoichiometric coefficients, molecular sizes and interaction parameters affect LLPS and aging. Finally, we analyze the thermodynamic stability of the two-phase system and identify the spinodal regions, and outline the generalization of our approach to reactive biomolecular systems with an arbitrary number of components.

\end{abstract}

\maketitle

\section{\label{sec:level1} Introduction}
Traditionally, intracellular organization of organelles has been associated with compartments that are surrounded by membranes. Modern imaging approaches indicate that membrane-less organelles exist outside of this classical view \cite{Brangwynne_2009}. These organelles are micron-sized clusters comprised of macromolecules such as RNAs and proteins, which can emerge via liquid-liquid phase separation (LLPS), forming droplets (``biomolecular condensates'') \cite{Hyman_2014_review}. More intriguingly, the dynamics of these initially liquid-like condensates often slows down over time, exhibiting viscoelastic or solid-like properties \cite{SHIN_science,Jawerth_Science}. This time-dependent liquid-to-solid-like transition (``aging'') is a characteristic behavior associated with several neurodegenerative diseases such as Alzheimer’s disease, Parkinson’s disease and amyotrophic lateral sclerosis (ALS) among others \cite{Jucker_Nature}. Therefore, the coupling behavior between LLPS and aging of condensates should not be overlooked.

The scope of LLPS can be expanded from binary solutions to various systems in which several species of macromolecules interact with each other. Bracha et al.~designed an oligomerizing biomimetic system (``Corelets'') to investigate the effects of multivalent interactions of intrinsically disordered protein regions (IDPs/IDRs) on LLPS \cite{bracha}. In their system, a spherical core with multiple photo-activatable domains recruit proteins when light is shone upon it. The new complex structure formed by the core and proteins behaves distinctly from its constituents, and it can in turn aggregate and phase separate from the buffer solution to form a liquid condensate. One can perceive this activation/deactivation process as a reversible light-induced chemical reaction. In fact, a number of experiments have already shown that macromolecules with different sizes and lengths, such as colloid-polymer solutions \cite{Poon_2002}, cholesterol and phospholipid systems \cite{RADHAKRISHNAN1999} and enzyme-protein mixtures \cite{rubisco, rubisco_He}, can readily react to form condensates. 
Theoretical works by Corrales and Wheeler \cite{Corrales_Wheeler} and Talanquer \cite{Talanquer} were the first to consider a reversible chemical reaction between the binary components of a liquid mixture to produce a third liquid component, which subsequently phase-separated from the reactant species. Radhakrishnan and McConnell \cite{RADHAKRISHNAN1999, RADHAKRISHNAN_PNAS} in turn extended this idea to the non-ideal mixing behavior of cholesterol, reactive and unreactive phospholipids, demonstrating that such models can be readily applied to systems of biological relevance. More recently, following the approach of Bazant \cite{Bazant_2013, Bazant_2017}, Kirschbaum and Zwicker formulated a thermodynamically consistent model to study chemical reactions in the context of biomolecular condensates while accounting for different molecular volumes of the reactants \cite{Kirschbaum_2021}. 

\begin{figure*}[!htb]
\includegraphics[scale=0.33]{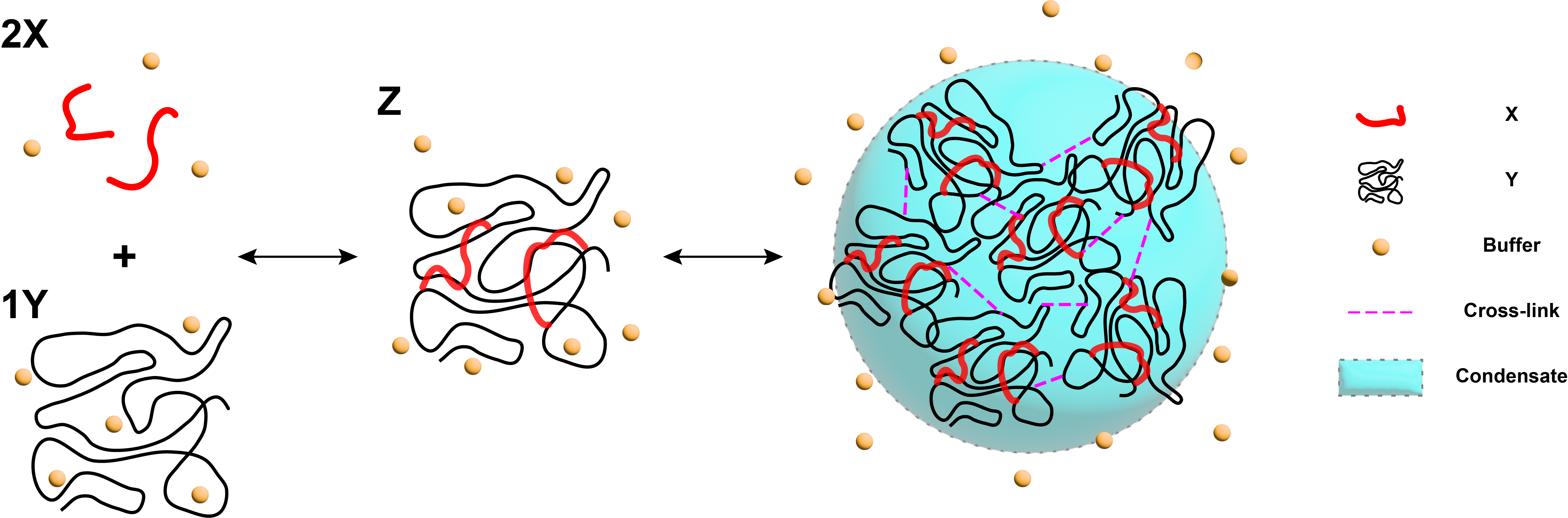}
\caption{\label{fig:1} Schematic of the reactive macromolecular mixture considered in this work. Two species, $X$ and $Y$, react to form a complex, $Z$, which subsequently phase separates from the buffer. The stoichiometric coefficients are set to $n = 2$ and $m = 1$ in this example.}
\end{figure*}

However, existing mesoscale theoretical models of multi-component LLPS or reaction-induced phase separation \cite{Mao_2019, Kirschbaum_2021} assume all species to be in a perfect liquid state, thus ignoring any aging processes which often have great physical significance in both polymer solutions and biomolecular systems. Aging can be caused by different types of microstructural changes, such as physical gelation and fibril formation. For example, gelatin-methanol-water mixture experiences phase separation and gelation \cite{TanakaT}; colloid-polymer solutions can undergo gelation \cite{Poon_2002}; and protein condensates show solid-like properties and various non-spherical morphologies when aging \cite{PATEL20151066, PESKETT2018588}. To partially address the role of aging on LLPS, Berry et al.~proposed a minimal mesoscale kinetic model for coupled ternary phase separation, gelation and chemical reactions \cite{Berry_2018}, but did not provide quantitative results; hence the need for a unified theory to address these problems. 

To remedy the aforementioned deficiencies in the existing theoretical approaches, we have developed a general framework to study the interplay between phase separation, chemical reactions and aging (gelation) operating concurrently in a multi-component macromolecular system. In this manuscript, we mainly focus on the formulation of the model, construction of phase diagrams, and identification of spinodal regions. We investigate how physical parameters such as stoichiometric coefficients, molecular sizes and interaction parameters affect LLPS and aging. A detailed study of the coupled diffusive kinetics, chemical reactions and aging behavior will be presented in a separate publication.  

\section{Model framework}
\begin{figure*}[!htb]
\centering
\includegraphics[scale=0.45]{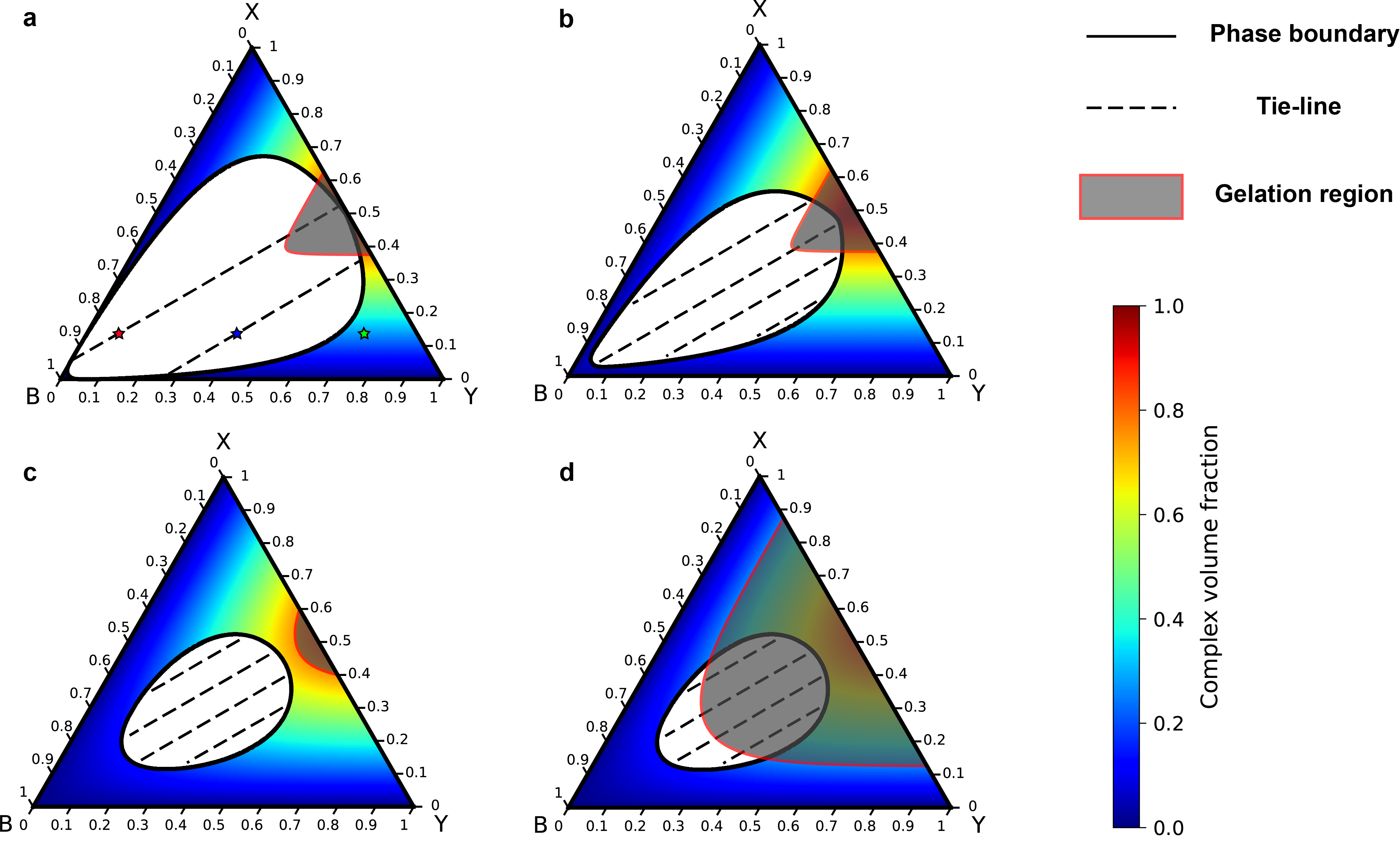}
\caption{\label{fig:2} Effects of varying interaction parameter, equilibrium constant and gelation criterion on phase behavior of the quaternary mixture at fixed stoichiometry ($n=m=1$), molecular volumes ($v_x=v_y=1$) and degrees of polymerization ($r_x=r_y=r_z=1$). The interaction parameter, equilibrium constant and gelation criterion were set to: (a) $\chi_{zb}=4$, $K=100$, $Z^\ast=0.6$ and $p=0.8$; (b) $\chi_{zb}=3$, $K=100$, $Z^\ast=0.6$ and $p=0.8$; (c) $\chi_{zb}=3$, $K=10$, $Z^\ast=0.6$ and $p=0.8$; (d) $\chi_{zb}=3$, $K=10$, $Z^\ast=0.2$ and $p=0.8$.}
\end{figure*}

We first consider an initially ternary system within which a reversible chemical reaction occurs between two molecular species $X$ and $Y$ immersed in a buffer solution ($B$); generalization of the model to systems with more components is briefly outlined in Section IV. Specifically, $X$ and $Y$ form a complex $X_nY_m$ ($Z$) via
\begin{equation}
\label{eqn:simple_reaction}
n X+m Y \leftrightarrow X_{\mathrm{n}} Y_{\mathrm{m}} \equiv Z,
\end{equation}
where $n$ and $m$ denote the stoichiometric coefficients. 
The four molecular species defined by their respective volume fractions, $X$, $Y$, $Z$ and $B$, form a liquid mixture at equilibrium. 
The mixture is taken to be incompressible such that $B=1-X-Y-Z$. 
Figure 1 illustrates a reaction between two molecules and one long polymer chain in forming a complex. 
The complexes subsequently aggregate to form a macroscopic condensate via LLPS. 

Inspired by the approach of Radhakrishnan and McConnell \cite{RADHAKRISHNAN1999}, we define a free energy density for the liquid mixture 
\begin{equation}
\label{eqn:fliquid}
\begin{aligned}
&f_{\mathrm{liquid}}(X,Y,Z) =\frac{X}{r_x} \ln X+ \frac{Y}{r_y} \ln Y+ \frac{Z}{r_z}\left(\ln Z + \mu_z\right) \\
&+ B\ln B +\chi _{xy}XY+\chi _{xz}XZ + \chi _{yz} YZ \\
&+ \chi _{xb}XB +\chi _{yb}YB + \chi _{zb}ZB,
\end{aligned}
\end{equation}
where $r_x$, $r_y$ and $r_z$ denote degrees of polymerization of $X$, $Y$ and $Z$.
In this model, all chemical potentials are constant and set equal to zero for convenience, with the exception of the chemical potential of the complex, $\mu_z=-\ln K$, where $K$ denotes the equilibrium constant. The interaction parameters $\chi_{ij}$ determine whether two different species ($i$ and $j$) attract or repel each other, which in turn control the global mixing/demixing behaviors.

Phase separation of polymers in solution may in addition induce strong associations in the form of cross-links or microcrystals between parts of the polymer chains, leading to the formation of a thermoreversible, physical gel \cite{TanakaT, TanakaF}. 
To incorporate gelation of the biomolecular mixtures simultaneously with phase separation, an order parameter $\phi \in [0, 1]$ is employed to quantify the gel concentration. 
The liquid-to-gel transition within the clusters of molecular complexes is captured using the simple free energy density \cite{Sciortino}
\begin{equation}
\label{eqn:fgel}
f_{\mathrm{gel}}(Z, \phi) = f_{\mathrm{g}}\left[-\frac{g(Z)}{2} \phi^{2}+\frac{\phi^{3}}{3}\right],
\end{equation}
where $f_g>0$ denotes a characteristic energy density scale, and the term $g\left(Z\right)$ couples gel concentration to the complex volume fraction via
\begin{equation}
\label{eqn:gcomplex}
g(Z)=\frac{pZ - Z ^ *}{1-Z ^*},
\end{equation}
where
\begin{equation}
p=\frac{\exp \left(\Delta F_c / k_{\mathrm{B}} T\right) }{1+\exp \left(\Delta F_c / k_{\mathrm{B}} T\right)}.
\end{equation}
The parameter $p$ denotes the fraction of the monomers in the polymer which are in the proper configuration to form cross-links, such that $pZ$ is the volume fraction of cross-links in the system. 
$\Delta F_c$ denotes the change in energy when forming a cross-link in the chain, and $k_\mathrm{B}$ and $T$ are Boltzmann constant and temperature, respectively. $Z^\ast$ in turn denotes the critical complex volume fraction necessary to form a gel. 
The form of $f_{\mathrm{gel}}\left(Z,\phi\right)$ ensures that gelation only occurs when the condition $pZ>Z^\ast$ is satisfied. 
Together, the total free energy of the system with total volume $V$ is thus written as
\begin{equation}
\begin{aligned}
\frac{F}{k_\mathrm{B}T}&= V \left[ f_{\mathrm{liquid}}(X,Y,Z) + f_{\mathrm{gel}}(Z,\phi)\right].
\label{eq:main}
\end{aligned}
\end{equation}
Equation (\ref{eq:main}) forms the starting point of the analysis of the coupled phase separation and aging behavior of the (effectively) quaternary system.

\section{Results}

\subsection{Ternary phase diagram}

\begin{figure*}[!htb]
\centering
\includegraphics[scale=0.45]{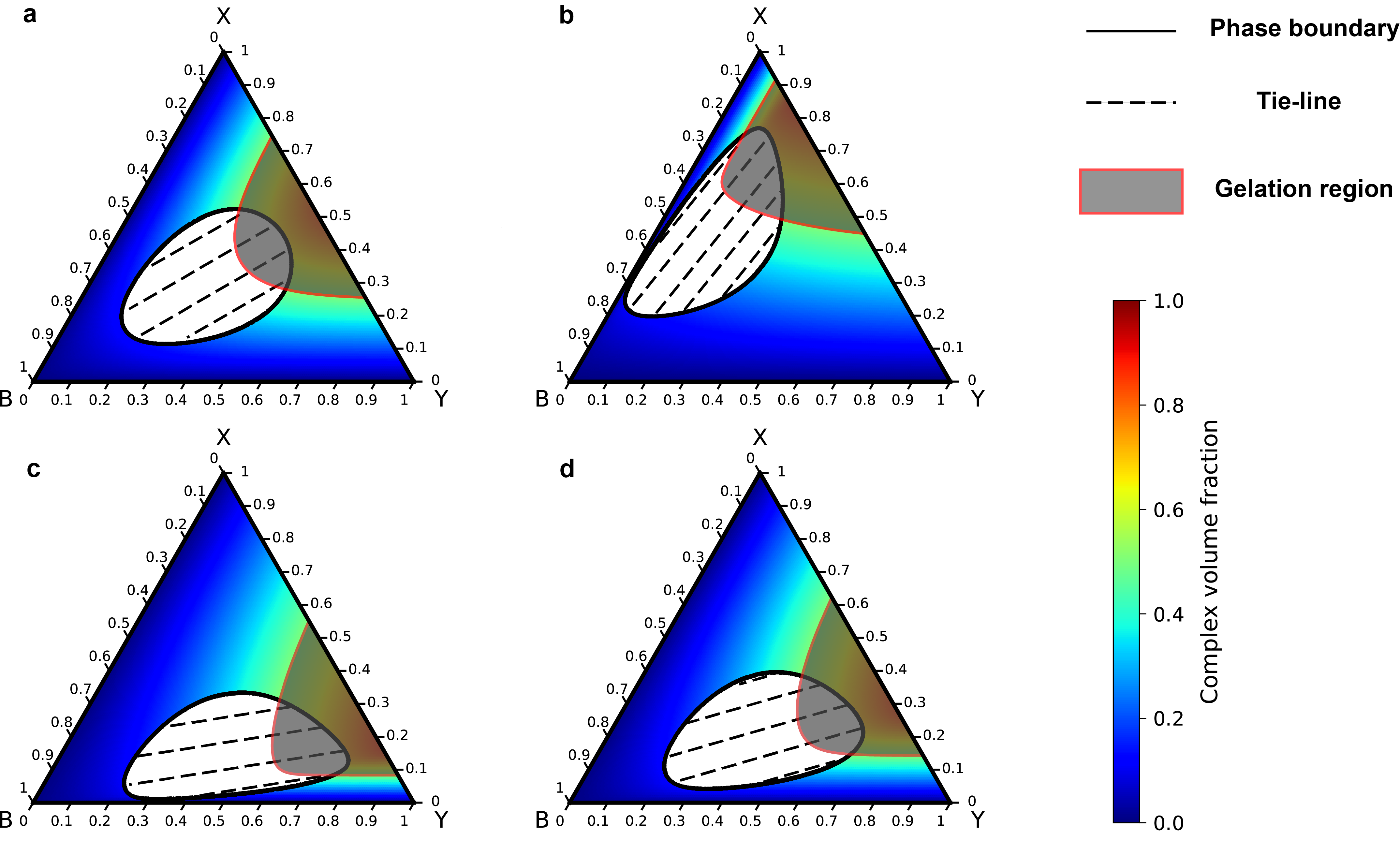}
\caption{\label{fig:3} Effects of varying stoichiometric coefficients of the reaction on phase behavior of the quaternary mixture at fixed interaction and gelation parameters ($\chi_{zb}=3$, $K=10$,$Z^\ast=0.2$ and $p=0.4$), molecular volumes ($v_x=v_y=1$) and degrees of polymerization ($r_x=r_y=r_z=1$). The stoichiometric coefficients were set to: (a)  $n=m=1$; (b) $n=5$, $m=1$; (c) $n=1$, $m=5$; (d) $n=2$, $m=5$.}
\end{figure*}

Phase diagrams are very useful in describing compositions and phase behaviors of a system, including phase coexistence and nature of phase transitions (e.g., continuous vs.~discontinuous). While multi-component phase diagrams have been widely used in the field of materials science, they rarely appear in the context of biological and chemically reactive systems. Veatch and Keller mapped phase boundaries of DPPC/DOPC/Chol mixture on a ternary phase diagram \cite{VEATCH_2003}. Radhakrishnan and McConnell in turn calculated the phase diagram for cholesterol and two phospholipids highlighting two-phase coexistence region \cite{RADHAKRISHNAN_PNAS}. More recently, Shin and Brangwynne proposed a hypothetical phase diagram for protein condensates, which incorporates phase coexistence and phase transitions between liquid, disordered ``glassy'' and solid/crystalline states \cite{SHIN_science}. 

To better understand the interplay between LLPS and gelation, we have constructed ternary phase diagrams of $X$, $Y$ and $B$ describing a reactive system that has no complex initially. The number fraction of the complex formed by chemical reaction, $\gamma$, serves as a reaction progress parameter \cite{RADHAKRISHNAN1999}. Then, the volume fraction of the complex formed can be expressed as $Z=v_z\gamma$, considering volume conservation during the reaction, $v_z=nv_x+mv_y$, where $v_x$, $v_y$ and $v_z$ denote molecular volumes of $X$, $Y$ and $Z$, respectively \cite{Kirschbaum_2021}. We can express $f_{\mathrm{liquid}}$ as a function of initial volume fractions for $X$ and $Y$, $x_0$, $y_0$, and reaction progress parameter with range $0\le\gamma\le\mathrm{min}\left(\frac{x_0}{nv_x},\ \frac{y_0}{mv_y}\right)$: 
\begin{equation}
\label{eqn:gamma-fliquid}
\begin{aligned}
&f_{\mathrm{liquid}}\left(x_0,y_0,\gamma\right)=\frac{\left(x_0-nv_x\gamma\right)}{r_x}\ln{\left(x_0-nv_x\gamma\right)}\\
&+\frac{\left(y_0-mv_y\gamma\right)}{r_y}\ln{\left(y_0-mv_y\gamma\right)}+\frac{v_z\gamma}{r_z}\left[\ln{\left(v_z\gamma\right)}-\ln{K}\right] \\
&+\left(1-x_0-y_0\right)\ln{\left(1-x_0-y_0\right)} 
+\chi_{yz}v_z\gamma\left(y_0-mv_y\gamma\right)\\
&+\chi_{xy}\left(x_0-nv_x\gamma\right)\left(y_0-mv_y\gamma\right)+\chi_{xz}v_z\gamma\left(x_0-nv_x\gamma\right)\\
&+ \left[ \chi_{xb}\left(x_0-nv_x\gamma\right) +\chi_{yb}\left(y_0-mv_y\gamma\right) \right]\left(1-x_0-y_0\right) \\
&+\chi_{zb}v_z\gamma\left(1-x_0-y_0\right).
\end{aligned}
\end{equation}
The computed phase diagrams are then constructed by first minimizing $f_{\mathrm{liquid}}\left(x_0,y_0,\gamma\right)$ with respect to $\gamma$ for given values of $x_0$ and $y_0$. The value of $\gamma$ obtained, $\gamma_\mathrm{min}$, then yields a free energy function $f_{\mathrm{liquid}}\left(x_0,y_0,\gamma_{\mathrm{min}}\right)$, from which the phase diagram can be calculated via the convex hull construction \cite{Wolff_2011,Mao_2019}. The algorithm not only determines phase coexistence regions, but also automatically generates tie-lines in those regions, thus determining the equilibrium compositions and complex volume factions. 

\begin{figure*}[!htb]
\centering
\includegraphics[scale=0.45]{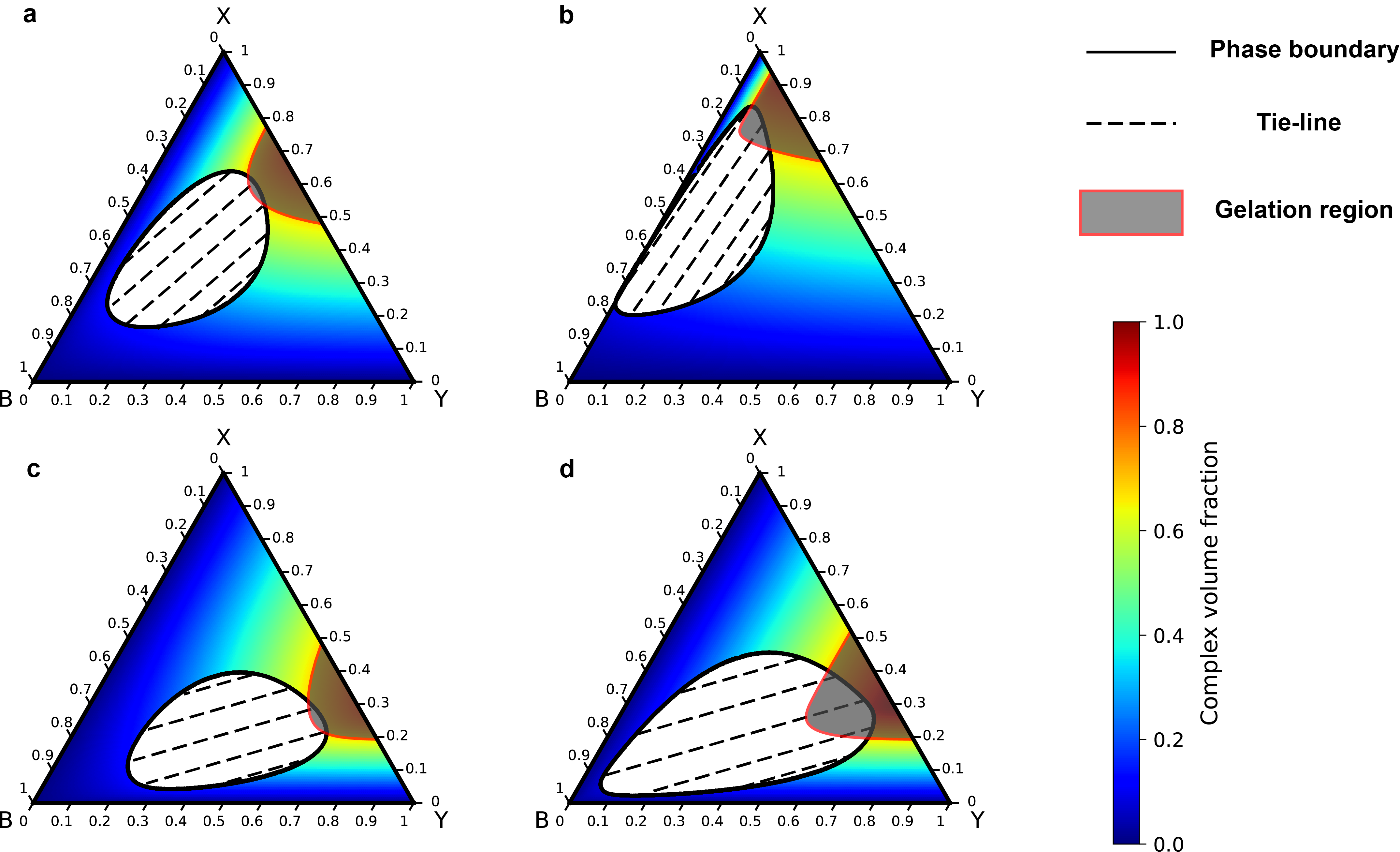}
\caption{\label{fig:4} Effects of varying molecular volumes and degrees of polymerization on phase behavior of the quaternary mixture at fixed stoichiometric coefficients ($n=2$, $m=1$) and interaction and gelation parameters ($\chi_{zb}=3$, $K=10$, $Z^\ast=0.4$ and $p=0.6$). The volumes and degrees of polymerization of each species were set to: (a)  $v_x=v_y=1$, $r_x=r_y=r_z=1$; (b) $v_x=5$, $v_y=1$, $r_x=r_y=r_z=1$; (c) $v_x=1$, $v_y=5$, $r_x=r_y=r_z=1$; (d) $v_x=1$,$v_y=5$, $r_x=r_z=1$, $r_y=5$.}
\end{figure*}

To illustrate how model parameters affect the phase behavior of the quaternary system, we next consider a simple case in which there only exists a repulsive interaction between $Z$ and $B$, controlled by a positive $\chi_{zb}$ parameter, while all other interaction parameters are set to zero. 
The phase diagrams shown in Fig.~2 describe symmetric systems with fixed $n=m=1$, $v_x=v_y=1$, $r_x=r_y=r_z=1$, but different values for $\chi_{zb}$, $K$, $Z^\ast$ and $p$. The white areas represent two-phase coexistence regions, in which the mixture will phase separate into a complex-poor and complex-rich phase as dictated by the tie-lines (dashed lines). 
Outside of the two-phase coexistence region is the single-phase region representing a homogeneous mixture of $X$, $Y$, $Z$ and $B$ at equilibrium colored by the equilibrium complex volume fraction. 
The gelation regions are shaded in grey for $Z>\frac{Z^\ast}{p}=Z_{gel}$; $Z_{gel}$ was set to $0.75$ in the phase diagrams in Figs.~2a, b and c and $0.25$ in Fig.~2d. In Fig.~2a, three choices for initial compositions $x_0$ and $y_0$, $(0.15,0.1)$, $(0.15,0.4)$ and $(0.15,0.75)$, are labelled as red, blue and green stars, respectively. 
Each of these systems yields a markedly different outcome: mixture initiated at red star will phase separate and undergo gelation in the condensed complex phase as the right end of the tie-line enters gelation region; 
mixture initiated at blue star will phase separate without experiencing any gelation; and initializing the system at the green star yields a homogeneous liquid solution with no gelation.

Upon decreasing the repulsive interaction between $Z$ and $B$, the two-phase coexistence region in Fig.~2b shrinks in the direction perpendicular to the tie-lines. 
Similarly, when the formation of the complex phase becomes less favorable chemically via a ten-fold decrease in $K$, the two-phase coexistence region moves further away from the $B$ corner, as shown in Fig.~2c. 
In addition, we observe that the two-phase coexistence and gelation regions no longer overlap, indicating that all phase separated domains will remain liquid-like. 
On the contrary, in Fig.~2d, with $Z_{gel} =0.25$, the entire right boundary of the two-phase coexistence region resides within the gelation region, indicating that all initial compositions inside the two-phase coexistence region lead to formation of a gel (either confined to droplets or system-spanning network as dictated by the volume fraction of $Z$). We note that there is also a distinct possibility of forming a gel network without phase separation if the initial composition is inside the single-phase gelation region.

The effects of stoichiometric coefficients on coupled phase separation and gelation behavior of the system were also investigated by systematically varying $n$ and $m$ in the reaction between two small molecule species, defined by the constants $v_x=v_y=1$, $r_x=r_y=r_z=1$. In addition, $p=0.4$ and $Z_{gel}=0.5$ were kept fixed in all the four cases shown in Fig.~3. 
Now, Fig.~3a shows a symmetric phase diagram with $n=m=1$. 
In Fig.~3b, we consider a different stoichiometric ratio for the reaction between $X$ and $Y$, namely $5X+Y \leftrightarrow Z$. 
The resulting phase diagram displays a markedly asymmetric two-phase coexistence region and complex volume fraction distribution.
This is simply due to fact that in forming the complex, the system consumes more $X$ than $Y$, and hence the phase diagram becomes skewed towards high $X$ volume fraction. Setting $n = 1$ and $m = 5$ naturally yields a phase diagram which is mirrored by the line $X=Y$ (Fig.~3c). 
Finally, changing the stoichiometric ratio to $n =2$, $m =5$ gives rise to a phase diagram (cf.~Fig.~3d) closer to the symmetric case, in agreement with intuition.

We have also studied the effects of different molecular volumes and degrees of polymerization on the phase behavior of the system. 
To this end, Fig.~4a describes two species with equal size and degree of polymerization as discussed in Figs.~2 and 3. 
The co-existence region is slightly skewed towards X-axis as we set $n = 2$, $m = 1$, $\chi_{zb}=3$, $K=10$, $Z^\ast=0.4$ and $p=0.6$, which are held constant for all four representative cases. 
If $X$ represents a larger ``blob'' than $Y$, e.g., $v_x =5$, $v_y=1$, $r_x=r_y=r_z=1$, the phase diagram again becomes strongly skewed towards high $X$ volume fractions (Fig.~4b). If we now reverse the volumes for $X$ and $Y$, i.e. $v_x =1$ and $v_y=5$, we observe that the phase diagram in Fig.~4c is identical to that in Fig.~3d except for the gelation region. This is due to the fact  $nv_x$ and $mv_y$ show up as products in Eq.~(\ref{eqn:gamma-fliquid}), resulting in the same free energy. 

Next we consider a more complex case where a long polymer $Y$ with volume $v_y=5$, reacts with two small $X$ molecules with volume $v_x = 1$. 
The polymer also has higher degree of polymerization than the small molecule, such that $r_x=1$, $r_y = 5$, and the complex has a granular structure with $r_z = 1$. 
The phase diagram in this case  (cf.~Fig.~4d) becomes more skewed towards the $Y$-axis and the two-phase coexistence region is significantly larger than the previous cases (while keeping $\chi_{zb}$ and $K$ fixed), implying that higher degrees of polymerization may facilitate LLPS and/or gelation at lower volume fractions of $X$ and/or $Y$.

\subsection{Spinodal behavior}
LLPS may proceed either via nucleation and growth or spinodal decomposition. 
In the spinodal region, the mixture becomes globally unstable towards small compositional fluctuations and results in spontaneous phase separation without nucleation. 
Identifying such regions is important for both numerical simulations and experiments.
Thus, in addition to the phase boundaries (binodal lines) displayed in the phase diagrams in Figs.~2, 3, and 4, we have also determined the spinodal regions via a standard quadratic approximation. 
That is, for a given initial composition $(x_0, y_0)$, we expand $f_{\mathrm{liquid}}$ from Eq.~(\ref{eqn:gamma-fliquid}) up to 2$^{nd}$ order in the compositional variations:
\begin{equation}
\begin{aligned}
f_{\mathrm{liquid}}(x, y) &\approx f_{\mathrm{liquid}}\left(x_{0}, y_{0}\right) + \vec{\nabla} f_{\mathrm{liquid}}\left(x_{0}, y_{0}\right) \cdot 
\begin{bmatrix}
x-x_{0} \\ y-y_{0}
\end{bmatrix} \\
&+\frac{1}{2}\begin{bmatrix} x-x_{0}, \  y-y_{0}\end{bmatrix} H_{f}\left(x_{0}, y_{0}\right)\begin{bmatrix} x-x_{0} \\ y-y_{0}\end{bmatrix},
\end{aligned}
\end{equation}
where $H_{f}\left(x_{0}, y_{0}\right)$ denotes the Hessian matrix ($2\times 2$) for $f_{\mathrm{liquid}}(x, y)$ evaluated at $\left(x_{0}, y_{0}\right)$. 
The spinodal region is identified as the one wherein at least one of the two eigenvalues of $H_f$ is negative, while both the binodal and one-phase regions will have two positive eigenvalues, as appropriate for a concave-up free energy landscape. 

\begin{figure}[!htb]
\includegraphics[scale=0.3]{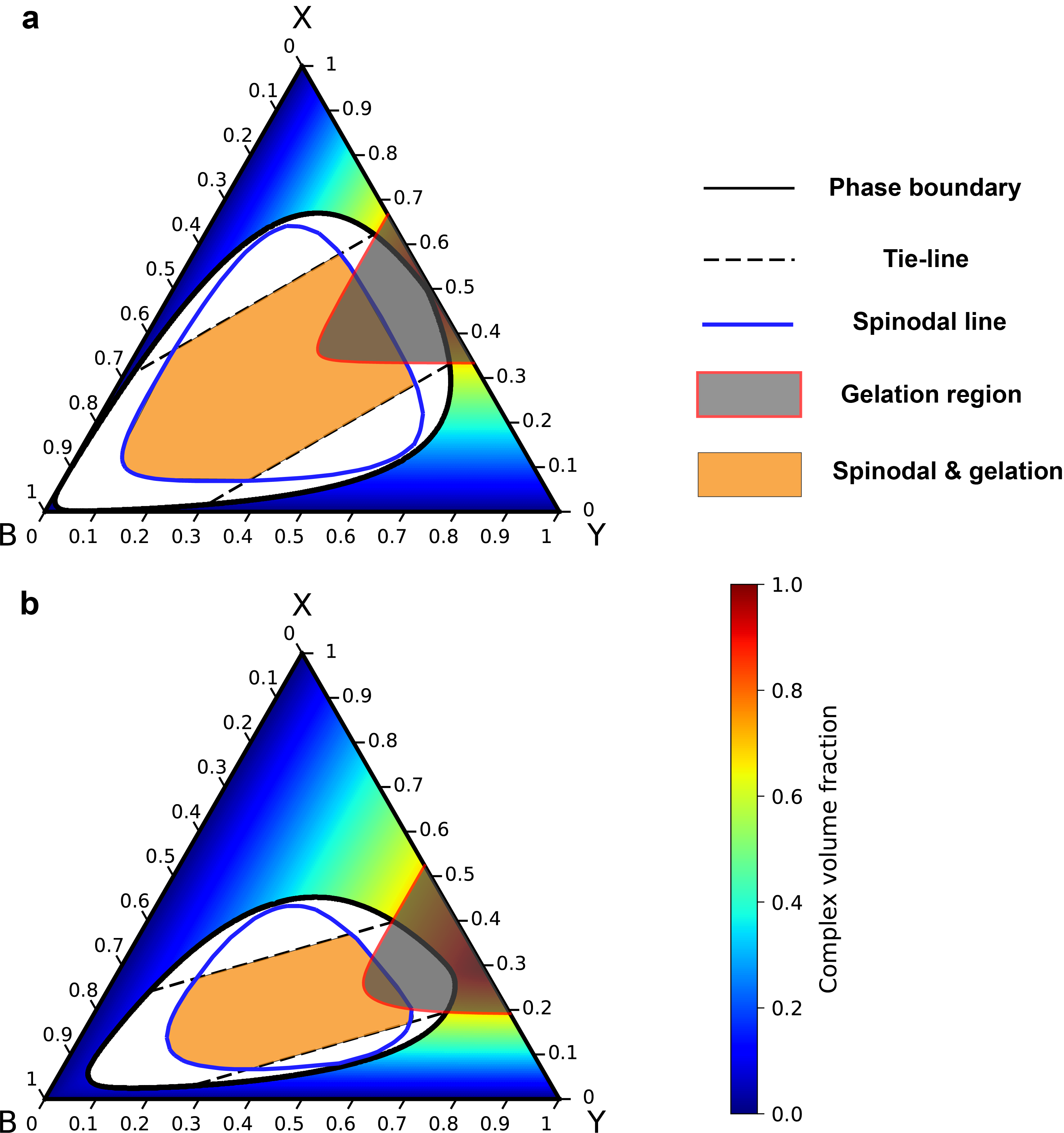}
\caption{\label{fig:5} Two representative phase diagrams with significantly overlapping spinodal and gelation regions. The parameter values employed in the construction of the phase diagrams were set to: (a) $n=m=1$, $v_x=v_y=1$, $r_x=r_y=r_z=1$, $\chi_{zb}=4$, $K=100$, $Z^\ast=0.4$ and $p=0.6$; and (b) $n=2$, $m=1$, $v_x=1$,$v_y=5$, $r_x=r_z=1$, $r_y=5$, $\chi_{zb}=3$, $K=10$, $Z^\ast=0.4$ and $p=0.6$. Even though the gelation regions make only small ``excursions'' within the two-phase coexistence regions, significant fractions of the initial compositions within the spinodals would lead to phase separated gel-like domains.}
\end{figure}

Having thus identified the thermodynamically unstable regions, we then delineate them in the phase diagrams as shown in Fig.~5.
As a consistency check, we note that the spinodal lines intersect the binodals only at the two critical points, for both symmetric (cf.~Fig.~5a) and asymmetric (cf.~Fig.~5b) cases. 
We further identify inside the spinodal regions the initial compositions where spinodal decomposition and gelation happen concurrently (colored as orange). It is noteworthy that even though the gelation regions make only small ``excursions'' within the two-phase coexistence regions, significant fractions of the initial compositions within the spinodals -- those easily triggered to display phase separation either numerically or experimentally -- would lead to phase separated gel-like domains.

\section{Conclusions}
In this work, we have formulated a thermodynamic model that captures chemical reactions, phase separation and gelation of macromolecular mixtures occurring concurrently. 
We have shown that chemical reactions may significantly alter the phase behavior when considering the effects of different stoichiometric coefficients, polymer sizes, degrees of polymerization, equilibrium constants and interaction strengths. Marked asymmetries in the phase diagrams were observed for systems in which the stoichiometric coefficients, molecular volumes and/or degrees of polymerization were significantly different between the two reactant molecular species. Furthermore, numerical identification of the spinodal regions demonstrated that in systems in which the gelation region overlaps with the two-phase coexistence one, large fractions of initial compositions within the spinodal regions would lead to phase separated gel-like domains. 

Generalization of our model to systems with even larger numbers of components is straightforward. To this end, consider an $N$-component system (where $N$ accounts for all molecular species present, including reactants, products and un-reactive ones) with volume fractions $\varphi_i$ subject to the incompressibility constraint $\sum _i ^N \varphi _i = 1$. Here the reactions are not restricted to binary ones, and we make the assumption that the reactants/products in one reaction do not react with other reactants/products. For $M$ reversible reactions, we define an $M \times N$ stoichiometric matrix $S_{ij}$. Equation (\ref{eqn:simple_reaction}) then generalizes to 
\begin{equation}
\sum_{j=1}^N S_{ij} \varphi_j  = 0 \,\,\,\,\,\, {\rm{for}} \,\,\,\,\,\, i=1,...,M,
\end{equation}
where the products (reactants) have positive (negative) entries for $S_{ij}$. Furthermore, let us define an $M \times N$ ``participation matrix'' $P_{ij}$ such that $P_{ij}=1$ if the $j^{th}$ component participates in the $i^{th}$ reaction while $P_{ij}=0$ otherwise.   Now, volume conservation is enforced via the $M$ constraints
\begin{equation}
\sum_{j=1}^N S_{ij} v_j  = 0 \,\,\,\,\,\, {\rm{for}} \,\,\,\,\,\, i=1,...,M.
\end{equation}
The free energy density in Eq.~(\ref{eqn:fliquid}) then generalizes to
\begin{equation}
f_\mathrm{liquid}\left(\left\{ \varphi _j \right\}\right) = \sum _{j=1} ^N \frac{\varphi_j}{r_j} \left( \ln \varphi _j + \mu _j \right) + \frac{1}{2}\sum _{j,k=1} ^N \chi _{jk} \varphi _j \varphi _k,
\end{equation}
where by convention $\chi _{jk} = 0$ for $j=k$. Once the stoichiometric matrix $S_{ij}$, the interaction matrix $\chi_{jk}$ and the chemical potentials $\mu_j$ have been specified, phase diagrams can be constructed by following the procedure as in the quaternary system. That is, Eq.~(\ref{eqn:gamma-fliquid}) can be generalized by introducing $M$ reaction progress parameters $\gamma _i$ for all distinct reactions:
\begin{equation}
\begin{aligned}
&f_\mathrm{liquid}\left(\left\{ \varphi _j ^0 \right\}, \left\{\gamma _i \right\} \right) \\
&= \sum_{i=1}^M \sum _{j=1} ^N P_{ij} \biggl\{ \frac{\varphi _j ^0 + S_{ij}v_j\gamma _i}{r_j} \Bigl[ \ln \bigl( \varphi _j ^0 + S_{ij}v_j\gamma _i \bigr)  + \mu _j \Bigr]  \\
& + \frac{1}{2} \sum _{j,k=1} ^N \chi _{jk} \left( \varphi _j ^0 + S_{ij}v_j\gamma _i\right)\times \left(\varphi _k ^0 + S_{ik}v_k\gamma _i\right) \biggr\}\\
& + \sum_{i=1}^M \sum _{j=1} ^N (1-P_{ij}) \Bigl[ \frac{\varphi _j ^0 }{r_j} \Bigl( \ln  \varphi _j ^0  + \mu _j \Bigr) + \frac{1}{2} \sum _{j,k=1} ^N \chi _{jk}  \varphi _j ^0 \varphi _k ^0 \Bigr],  
\end{aligned}
\end{equation}
where $\varphi _j^0$ denotes the volume fraction of species $j$ before mixing, thus the volume fractions of the products are initially zero. Subsequently, $\gamma _{i,\mathrm{min}}$ can be computed by minimizing $f_\mathrm{liquid}\left(\left\{ \varphi _j ^0 \right\}, \left\{\gamma _i \right\} \right)$ with respect to the $\gamma _i$ with ease as the reactions are non-interfering with one another. The phase diagram can be calculated using the same procedures as discussed in Section III. Although the convex hull construction is conceptually straightforward, it is computationally challenging for higher-dimensional systems, e.g., $N>6$ \cite{Mao_2019}. In addition, assuming any species in the mixture can form a gel $\phi _j$ with a distinct microstructure, Eq.~(\ref{eqn:fgel}) and Eq.~(\ref{eqn:gcomplex}) can be generalized to
\begin{equation}
f_{\mathrm{gel}}(\left\{\varphi _j, \phi _j \right\}) = \sum _{j=1} ^N f_{\mathrm{g}}^j \left[-\frac{g\left(\varphi _j \right)}{2} \phi _j^{2}+\frac{\phi _j^{3}}{3}\right]
\end{equation}
and
\begin{equation}
g\left(\left\{\varphi _j \right\}\right) = \frac{p_j \varphi _j - \varphi _j ^*}{1-\varphi _j ^*},
\end{equation}
respectively.

Finally, it is important to stress that for quantifying the full non-equilibrium behavior of phase-separating systems which may or may not display aging, phase diagrams alone will not suffice; one has to resolve the full spatio-temporal dynamics of the molecular species and their aging behavior. To this end,  we have implemented a thermodynamically consistent  formulation of the dynamics, derived from an extension of Eq.~(\ref{eqn:fliquid}) to spatially-varying volume fractions and supplanted with appropriate mass conservation laws and reaction kinetics. This allows us to further generalize the $N$-component system to more complex scenarios in which reactants/products are not restricted to only one reaction. 
The detailed results including numerical simulations, the interplay between kinetics and morphology will be presented in a separate manuscript currently in preparation for submission.

\section*{ACKNOWLEDGMENTS} RZ and MPH were supported by the National Science Foundation (NSF) Materials Research Science and Engineering Center Program through the Princeton Center for Complex Materials (PCCM) (DMR-2011750). MS acknowledges the partial support of the National Science Foundation of China (NSFC) under the grant number 12272005.

\end{document}